\documentclass[runningheads]{llncs}

\author{
David E.~Bernal\inst{1,4,5} \and
Kyle E. C.~Booth\inst{2} \and Raouf~Dridi\inst{3} \and
Hedayat~Alghassi\inst{3} \and Sridhar~Tayur\inst{3} \and Davide~Venturelli\inst{4,5}
}

\institute{Department of Chemical Engineering, Carnegie Mellon University, Pittsurgh PA 15213, USA \\ \email{bernalde@cmu.edu} \and
Department of Mechanical \& Industrial Engineering, University of Toronto, Toronto ON M5S 3G8, Canada \\ \email{kbooth@mie.utoronto.ca} \and
Tepper School of Business, Carnegie Mellon University, Pittsburgh PA 15213, USA \\ \email{\{rdridi,halghass,stayur\}@andrew.cmu.edu} \and
Quantum AI Lab (QuAIL),  NASA Ames Research Center, Moffett Field {CA 94035,}~USA \and
USRA (RIACS), Mountain View CA 94043, USA
\\ \email{davide.venturelli@nasa.gov}
}

\authorrunning{D.E. Bernal et al.}

\usepackage{geometry}
\usepackage[utf8]{inputenc} 
\usepackage[T1]{fontenc}    
\usepackage{hyperref}       
\usepackage{url}            
\usepackage{booktabs}       
\usepackage{amsfonts}       
\usepackage{amsmath}
\usepackage{nicefrac}       
\usepackage{microtype}      
\usepackage{lipsum}
\usepackage{lscape}
\usepackage[numbers,sort&compress]{natbib}
\usepackage{cleveref}
\usepackage{threeparttable}
\usepackage{epsfig}
\usepackage{color}
\usepackage[export]{adjustbox}
\usepackage{subfig, color}
\usepackage{multirow}
\usepackage{floatrow}
\usepackage{pdflscape}
\usepackage{graphicx}
\usepackage{subfig}
\usepackage{soul}
\usepackage[normalem]{ulem}

\usepackage[svgnames]{xcolor}
\usepackage{tikz}
\usetikzlibrary{decorations.markings}
\usetikzlibrary{shapes.geometric}

\newcommand\nodescale{1.25} 

\pgfdeclarelayer{edgelayer}
\pgfdeclarelayer{nodelayer}
\pgfsetlayers{edgelayer,nodelayer,main}

\tikzstyle{none}=[inner sep=0pt]

\tikzstyle{rn}=[circle,fill=Red,draw=Black,line width=0.8 pt]
\tikzstyle{gn}=[circle,fill=Lime,draw=Black,line width=0.8 pt]
\tikzstyle{yn}=[circle,fill=Yellow,draw=Black,line width=0.8 pt]
\tikzstyle{auxiliary_qubit}=[circle,fill=Red,draw=none,scale=\nodescale,inner sep=3.5 pt, outer sep=-1pt]
\tikzstyle{logical_qubit}=[circle,fill=Black,draw=none,scale=\nodescale,inner sep=3.5 pt, outer sep=-1pt]
\tikzstyle{emb_logical_qubit}=[circle,fill=Black,draw=none,scale=\nodescale,inner sep=3.5 pt, outer sep=-1pt]
\tikzstyle{emb_auxiliary_qubit}=[circle,fill=Red,draw=none,scale=\nodescale,inner sep=3.5 pt, outer sep=-1pt]
\tikzstyle{unused_qubit}=[circle,fill=Gray,draw=none,scale=\nodescale,inner sep=3.5 pt, outer sep=-1pt]
\tikzstyle{arrow_end}=[circle,fill=none,draw=none,scale=.1]

\tikzstyle{simple}=[-,draw=Black,line width=1.000]
\tikzstyle{added}=[-,draw=Black,line width=1.000]
\definecolor{tempcolor}{rgb}{.9,.9,.9}
\tikzstyle{unused}=[-,draw=tempcolor,line width=0.500]
\tikzstyle{unused_added}=[-,draw=tempcolor,draw opacity=1,line width=0.5]
\tikzstyle{embedding}=[-,draw=Black,line width=3.250]
\tikzstyle{embedding_aux}=[-,draw=Red,line width=3.250]
\tikzstyle{arrow}=[-,draw=Black,postaction={decorate},decoration={markings,mark=at position .5 with {\arrow{>}}},line width=2.000]

\title{Integer programming techniques for minor-embedding in quantum annealers}

\date{December 2019}

\begin{document}

\maketitle
\begin{abstract}
A major limitation of current generations of quantum annealers is the sparse connectivity of manufactured qubits in the hardware graph.
This technological limitation has generated considerable interest, motivating efforts to design efficient and adroit minor-embedding procedures that bypass sparsity constraints. In this paper, starting from a previous equational formulation by Dridi et al. (arXiv:1810.01440), we propose integer programming (IP) techniques for solving the minor-embedding problem.
The first approach involves a direct translation from the previous equational formulation to IP, while the second decomposes the problem into an assignment master problem and fiber condition checking subproblems.
The proposed methods are able to detect instance infeasibility and provide bounds on solution quality, capabilities not offered by currently employed heuristic methods.
We demonstrate the efficacy of our methods with an extensive computational assessment involving three different families of random graphs of varying sizes and densities.
The direct translation as a monolithic IP model can be solved with existing commercial solvers yielding valid minor-embeddings, however, is outperformed overall by the decomposition approach. Our results demonstrate the promise of our methods for the studied benchmarks, highlighting the advantages of using IP technology for minor-embedding problems. 
 
\end{abstract}
 

\section{Introduction}

Graph minor theory (GMT), the central theme of this work, is prominent across many fields. In quantum computing, GMT is employed to extend the scope of problems that can be represented on current quantum annealing hardware \cite{kaminsky2004scalable, Choi2011}. Mapping a dense problem (logical) graph $Y$ to a sparse (target) graph $X$ can be achieved by
constructing connected subgraphs of the target graph $X$ from the high degree logical vertices $y$. The resulting mapping is called a minor-embedding of $Y$ inside $X$.  

Numerous heuristics for finding minor-embeddings have been proposed \cite{cai2014practical,bian2014discrete,yang2016graph}.
While these approaches are generally fast, they do not provide guarantees on the quality of the produced minor-embeddings nor can they prove the nonexistence of a minor-embedding for infeasible problems. An approach that attempts to address these shortcomings was recently introduced in \citet{dridi2018novel}. This approach uses tools from algebraic geometry and produces an equational formulation (as opposed to a purely combinatorial approach)  to the minor-embedding problem.  

In this paper, starting from this equational formulation, we propose integer programming (IP) techniques for tackling the embedding problem. Our proposed approaches differ from the computationally demanding Groebner bases computation used previously and are aimed at more efficiently computing embeddings while retaining the interesting properties that arise from the equational formulation of the problem.
Our first approach, detailed in Section \ref{sec:IP-constraints}, directly translates the previous equational formulation to IP, while our second approach decomposes the problem into an assignment master problem and fiber condition checking subproblems, as described in Section \ref{sec:const_gen}.
The proposed methods are able to detect instance infeasibility and provide bounds on solution quality, capabilities not offered by currently employed heuristic methods.
While recent work uses an approach with integer programming to address the embedding problem based on templates specific to D-Wave quantum annealers \cite{Serra2019}, the techniques we present in this paper are hardware agnostic. 

We conduct an extensive empirical analysis involving a benchmark consisting of three different families of random graphs in Section \ref{sec:results}.
There we present our results on an illustrative and challenging case for heuristics, which motivates the use of IP over Computational Algebraic Geometry (CAG) methods in random structured and unstructured graphs, and in applications for quantum annealing.
The results of the experiments indicate that, while the IP-based methods are slower than currently employed heuristics whenever the heuristics are able to find an embedding, the IP methods provide infeasibility proofs and quality guarantees which the heuristics are unable to provide.
Furthermore, comparing the monolithic IP against the decomposition, our experiments suggest that the decomposition results in a more efficient way to find concise embeddings.
Depending on the tested instances the decomposition approach does not perform as efficiently as the monolithic IP approach in providing optimality or infeasibility guarantees, especially seen in the illustrative example and small structured graphs. 
We provide concluding remarks in Section \ref{sec:conclusions}.

\emph{Notations. }
All graphs considered in this paper are simple and undirected. 
We use ${\bf V}(X)$ and ${\bf E}(X)$ to denote the vertex and edge sets of a graph $X$, respectively. We also define $n = |{\bf V}(X)|$, and $m = |{\bf V}(Y)|$. Finally, given a vector $v$, $\mathbf{v}$ denotes the 
concatenation  $\mathbf{v}=(v_1,\cdots,v_{|\mathbf{v}|})$. 

\section{The equational model for embedding} \label{sec:equational}
Let $X$ be a fixed target graph.
A \emph{minor-embedding} of the graph $Y$ inside $X$, is a map 
$\phi$ from ${\bf V}(Y)$ to the set of
connected subtrees of $X$,
that satisfies the following condition: for each $( y_1, y_2)\in {\bf E}(Y)$, there exists at least one edge in ${\bf E}(X)$ connecting the two subtrees $\phi(y_1)$ and $\phi(y_2)$.   
The condition that each {\it vertex model} $\phi(y)$ is a connected subtree of $X$ can be relaxed into $\phi(y)$ is a connected subgraph.
In the literature, there is another but equivalent definition of minor-embedding in terms of deleting and collapsing the edges of $X$.
This follows from the fact that, given a minor-embedding $\phi$, the graph $Y$ can be recovered from $X$ by collapsing each set $\phi(y)$ (into the vertex $y$) and ignoring (deleting) all vertices of $X$ that are not part of any of the subtrees~$\phi(y)$. 
For the sake of a simple and clean terminology, we shall use the term embedding instead of minor-embedding throughout the remainder of the paper. 
Suppose $\phi$ is an embedding of the graph $Y$ inside the graph $X$.
The subgraph of $X$ given by 
$
    \phi(Y) := \cup_{y\in {\bf V}(Y)}\phi(y)
$
is called a $Y$ minor in GMT.
In the context of quantum computations, it represents what the quantum processor sees since it does not distinguish between qubits representing different nodes of the logical graph or qubits representing the same node in the logical graph, {\it fibers}.
In practice, quantum annealers use a strong ferromagnetic coefficient to enforce these replicated values to be equal, i.e., acting as a single qubit.

\begin{figure}[h]
    \centering
    \subfloat  
    {{\includegraphics[width=3cm]{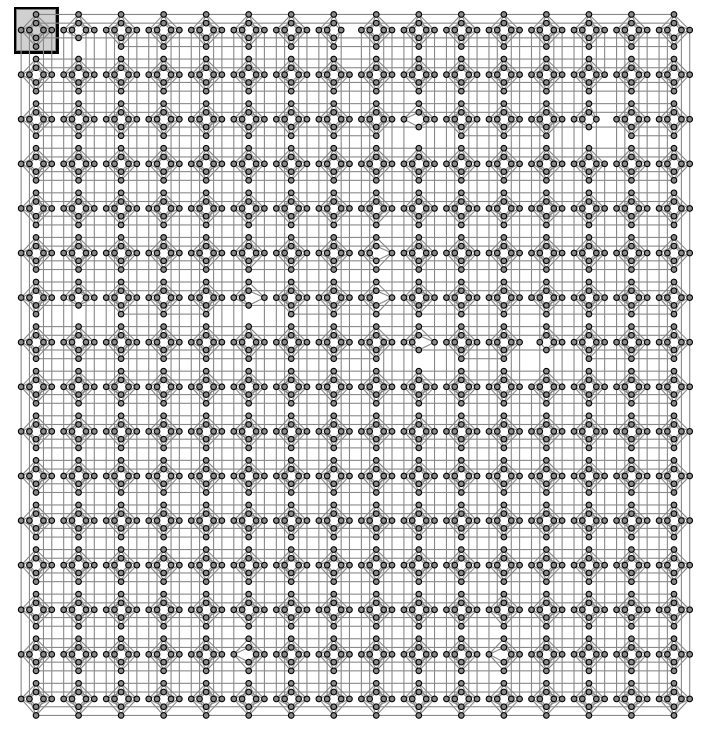} }}%
    \quad
    \centering
    \subfloat  
    {{\includegraphics[width=5cm]{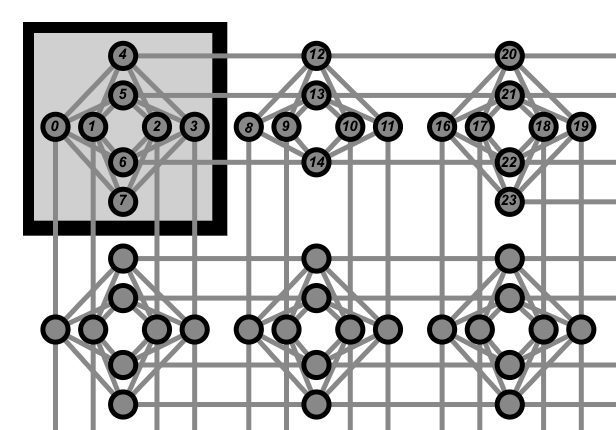} }}%
    \quad
    \centering
    \subfloat  
    {{\includegraphics[width=3cm]{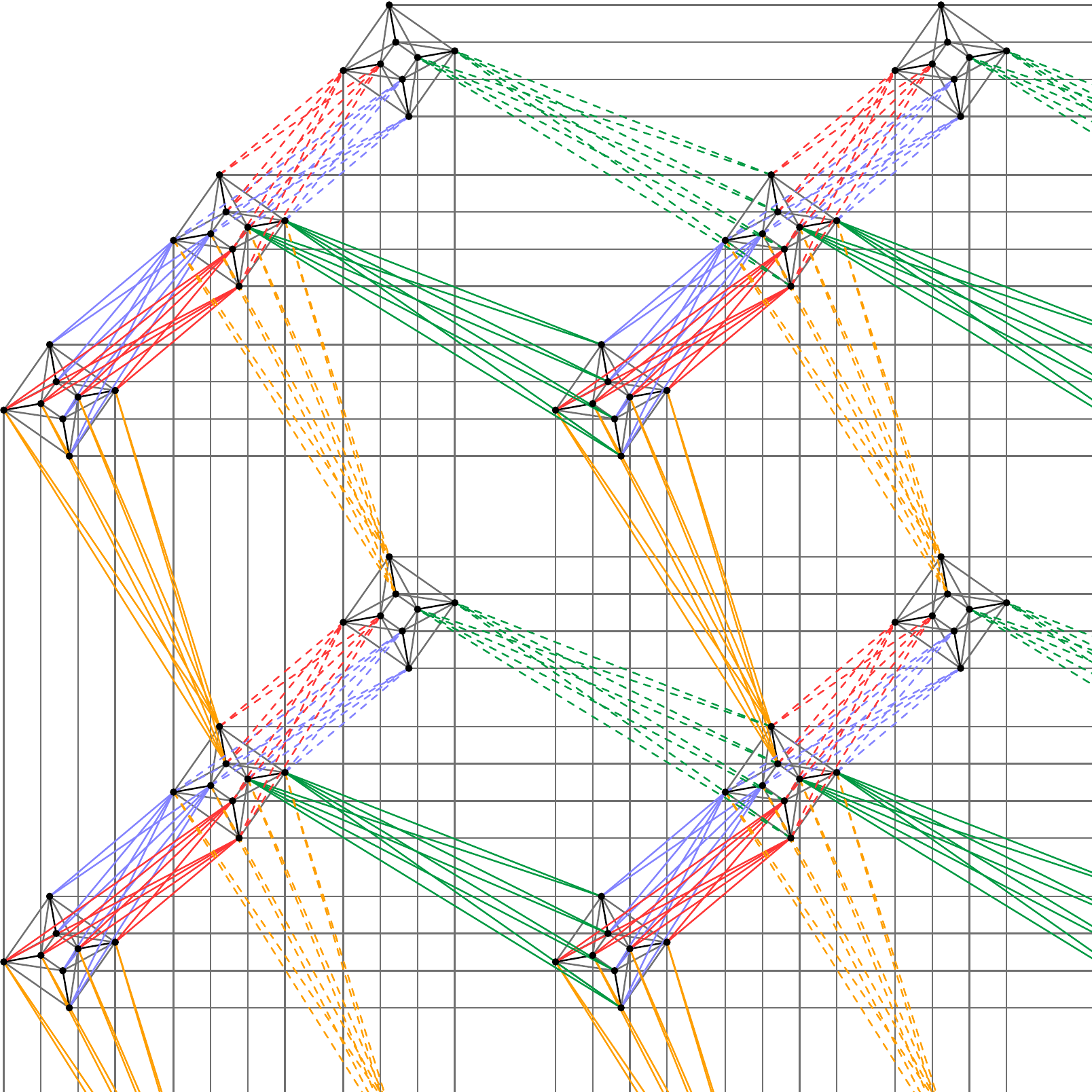} }}%
    \caption{{\small Cross representation of D-Wave Systems 2000Q processor working graph corresponding to an incomplete Chimera graph $\mathcal{C}_{4,16,16}$ (left and center) and Pegasus graph $\mathcal{P}_{4,2,2,3}$ (right). 
    }}
    \label{fig:Chimera}
  \end{figure}

The hardware configuration of the D-Wave quantum annealers is described as a graph known as Chimera graph.
The Chimera graph, $\mathcal{C}_{L,M,N}$, is a grid of $M \times N$ cells of $K_{L,L}$ biclique graphs connected in a nearest-neighbor fashion by means of non-planar edges \cite{neven2009nips}.
Figure~\ref{fig:Chimera} presents the working graph of the D-Wave 2000Q Quantum Annealer located at NASA Ames Research Center.  Specific heuristics for finding embedding inside Chimera graphs were developed in \cite{yang2016graph,teramoto2018graph,goodrich2018optimizing,okada2019improving} besides of the general embedding heuristics referenced above.
A new generation of quantum annealers is in development, the D-Wave {\em{Advantage}}, using the \emph{Pegasus} topology with increased connectivity~\citep{boothby2019next} (Figure~\ref{fig:Chimera}, right). 
The \emph{Pegasus} graph, $\mathcal{P}_{L,M,N,O}$, is composed of $O$ layers of $M \times N$ grids of $K_{L,L}$ biclique graphs with additional edges within  and among cells.
The \emph{Pegasus} topology has the number of layers fixed, $O=3$, with $K_{4,4}$ cells, i.e. $L=4$, with 4 additional edges each.

In the equational approach, embedding the logical graph $Y$ inside the target graph $X$, is represented by a surjective map  $\pi:X\rightarrow Y$, which  goes in the opposite direction of the map  $\phi:Y\rightarrow X$, introduced earlier such that: $\pi^{-1}(y)= \phi(y).$ The map $\pi$ is required to be surjective to guarantee that all logical qubits are embedded.
In geometry, the subgraph $\pi^{-1}(y)$ is called the {\it fiber} at $y$ of the {\it projection} $\pi$, and the mapping $\pi:X\rightarrow Y$ is a {\it fiber-bundle}.    
We can write:
\begin{equation}\label{pidef}
    \pi(x_i)= \sum_{j: y_j \in \textbf{V}(Y)} \alpha_{ij} y_j, \quad \forall x_i\in {\bf V}(X)
\end{equation}
where $\alpha_{ij}$ are binary coefficients. 
For this map to be well-defined we impose: 
\begin{equation}
\label{well_defined}
   \sum_{j: y_j \in \textbf{V}(Y)}  \alpha_{ij}\leq 1 \quad \forall x_i \in \textbf{V}(X), 
\end{equation}
that is, at most one $\alpha_{ij}$ is non-zero for each vertex in the graph $X$.
The unique non zero $\alpha_{ij}$ (if any) represents whether the physical qubit $x_i$ embeds $y_j$, i.e., $\phi(y_j)=x_i$.
When all the coefficients $\alpha_{ij}$ are zero, we get $\pi(x_i)=0$ indicating that the physical qubit is not used.
In other words, while the domain of definition of $\pi$ is ${\bf V}(X)$, its support is only a subset of ${\bf V}(X)$.
The other conditions included in the definition of the embedding $\phi$ (e.g., the connectivity of the fibers) and the desired properties of such an embedding (e.g., the size of the fibers) can similarly be translated into equational form.
\section{IP reformulation of polynomial equations}
\label{sec:IP-constraints}
We tackle the problem of determining the mapping $\pi$ using integer programming (IP).
IP is a mathematical optimization technique used for problems modeled as a set of decision variables taking on integer values, constrained by linear constraints and looking to optimize a linear objective function. The standard solution approach to IP models is branch-and-bound tree search. Indeed, due to their many practical applications, the computational capabilities of modern IP solvers have increased tremendously in recent years~\cite{Bixby2012}. These IP solvers are capable of proving instance infeasibility, and providing certificates of optimality and bounds on solution quality.

In our first approach, the previously proposed polynomial equations \cite{dridi2018novel} are reformulated such that they represent the original logic and are representable in the IP formalism (i.e., linear constraints involving integer variables). 


\subsection{Constraints}

Consider the mapping $\pi$ given by Eqn.~\eqref{pidef}. In this section, we present the IP formulation of the polynomial conditions, from \citet{dridi2018novel}, that the coefficients $\alpha_{ij} \in \{ 0,1 \} \quad \forall x_i \in {\bf V}(X), \forall y_j \in {\bf V}(Y)$ need to satisfy for $\pi$ to be a valid embedding. This constitutes the first contribution of the paper. Note that, with a slight abuse of notation, our IP approaches redefine $\alpha_{ij}$ as a binary decision variable equal to 1 if $x_i$ belongs to the vertex model of $y_j$, and 0 otherwise.

\noindent\emph{1. Minimum and maximum size.}
These constraints ensure that the total number of qubits is bounded within the number of variables in the original problem and the total number of qubits $n$.
 
\begin{equation}
    m \leq \sum_{i:x_i\in {\bf V}(X)} \sum_{j:y_j \in {\bf V}(Y)} \alpha_{ij} \leq n.
    \label{totalsize_ip}
\end{equation}


\noindent\emph{2. Well-definition of the map $\pi$.} This is captured by Eq.~\eqref{well_defined}.

\noindent\emph{3. Fiber size constraint.}
This constraint on the size of the vertex models $|\phi(y_j)|$, known as \textit{fiber size}, is given by:
\begin{equation}
     1 \leq \sum_{i:x_i \in {\bf V}(X)} \alpha_{ij} \leq k \quad \forall y_j \in {\bf V}(Y).
    \label{size_ip}
\end{equation}
where $k$ is the desired maximum size of each fiber $\pi^{-1}(y_j)$.
The lower bound  ensures that all the logical variables are embedded i.e., the map $\pi$ is a surjection on the set  ${\bf V}(X)$.
We also include the following constraint:
\begin{equation}
    1 \geq \alpha_{i_1j} + \alpha_{i_2j} \quad \forall x_{i_1}, x_{i_2} \in {\bf V}(X), \min d(x_{i_1}, x_{i_2}) > k, \forall y_j \in {\bf V}(Y). 
    \label{sizepath_ip}
\end{equation}
This additional refinement excludes pairs $x_{i_1}$ and  $x_{i_2}$ from being in the fiber~$\pi^{-1}(y_j)$ whenever their distance, $d(x_{i_1}, x_{i_2})$, is larger than $k$, the desired maximum size of the fiber.  
 
\noindent\emph{4. Fiber condition.} We require that 
each fiber to be a connected subtree of $X$:
\begin{equation}
\begin{aligned}
   \forall x_{i_1},x_{i_2} \in \pi^{-1}(y_j): \alpha_{i_1j} + \alpha_{i_2j} + \left( \sum_{c_k(x_{i_1},x_{i_2}) \in C_k(x_{i_1},x_{i_2})} \left(\gamma_{c_k,j} \right) - 1\right) \leq 2. 
    \label{fiber_ip}
\end{aligned}
\end{equation}
The binary variable $\gamma_{c_k,j}$ takes a value of 1 if a fiber $c_k(x_{i_1}, x_{i_2})$ is used in the vertex model of $y_j$, and 0 otherwise.
Here $c_k(x_{i_1}, x_{i_2})$ is a fiber of size $\leq k$ connecting the two physical qubits $x_{i_1}$  and $x_{i_2}$, and $\text{int}(c_k(x_{i_1},x_{i_2}))= c_k(x_{i_1}, x_{i_2})\backslash \{x_{i_1}, x_{i_2} \}$.
We also write $C_k(x_{i_1}, x_{i_2})$ to denote the set of all fibers of size $\leq k$  connecting $x_{i_1}$  and $x_{i_2}$. This condition  implies the existence of a unique fiber connecting the pair and completely contained in $\pi^{-1}(y_j)$.  This automatically  implies that $\pi^{-1}(y_j)$ is connected.
The binary  $\gamma_{c_k,j}$ can be defined using the following IP representable constraints: for all $c_k(x_{i_1},x_{i_2}) \in C_k(x_{i_1},x_{i_2})$ and for all $y_j \in {\bf V}(Y)$: 
\begin{equation} \label{gamma_eqs}
    \begin{aligned}
    \gamma_{c_k,j} &= \prod_{\ell:x_\ell \in \text{int}(c_k(x_{i_1},x_{i_2}))} \alpha_{\ell j} \Leftrightarrow 
    \begin{cases}
        \gamma_{c_k,j} & \leq \alpha_{\ell j} \quad \forall x_\ell \in \text{int}(c_k(x_{i_1},x_{i_2})) \\
        \gamma_{c_k,j} & \geq 1 - (k-1) + \sum_{\ell: x_\ell \in \text{int}(c_k(x_{i_1},x_{i_2}))} \alpha_{\ell j}
    \end{cases}
    \end{aligned}
\end{equation}
%
The constraint in Eq.~\eqref{fiber_ip} does not exclude the cases where 2 variables in the source graph $(y_{j_{1}},y_{j_{2}}) \in {\bf E}(Y)$ are mapped to 4 different qubits in a fiber $\{ x_{i_1}, \cdots, x_{i_4} \}$,  where the vertex models are intercalated, i.e. $\phi(y_{j_{1}}) = \{ x_{i_{1}},x_{i_{3}} \}, \phi(y_{j_{2}}) = \{ x_{i_{2}},x_{i_{4}} \}$.
The following constraint ensures that if two nodes in the target graph are in the vertex model of the same logical variable, and are not neighbors in the target graph, then one of the fibers joining them has to be active. 
\begin{equation}
\begin{aligned}
    \alpha_{i_{1}j} + \alpha_{i_{2}j} &- \sum_{c_k(x_{i_1},x_{i_2}) \in C_k(x_{i_1},x_{i_2})} \left( \gamma_{c_k,j} \right) \leq 1 \quad 
    \forall y_j \in {\bf V}(Y)\\
    \forall (x_{i_{1}},x_{i_{2}}) &\in {\bf V}(X), (x_{i_{1}},x_{i_{2}}) \not\in {\bf E}(X), \min d(x_{i_1}, x_{i_2}) \leq k
\end{aligned}
\label{fiber2_ip}
\end{equation}

\noindent\emph{5. Pullback condition.} We require that for each edge $(y_{j_1}, y_{i_2})$ in ${\bf E}(Y)$, there exists at least one edge	in ${\bf E}(X)$ connecting the fibers $\pi^{-1}(y_{j_1})$ and $\pi^{-1}(y_{i_2})$.
The way we guarantee this is by requiring that the quadratic form of the logical graph $y$ vanishes modulo the (pullback  along $\pi$ of the) quadratic form of the graph $X$. The details of this are in \citep{dridi2018novel}.
The resulting constraint can be written as
\begin{equation}
    1 \leq \sum_{i_1,i_2: (x_{i_1},x_{i_2}) \in {\bf E}(X)} \left( \delta^{\parallel}_{i_{1}i_{2}j_{1}j_{2}} + \delta^{\bot}_{i_{1}i_{2}j_{1}j_{2}} \right) \quad \forall (y_{j_1},y_{j_2}) \in {\bf E}(Y),
    \label{pullback_ip}
\end{equation}
\noindent where we have  introduced the binaries   $\delta^{\parallel}_{i_{1}i_{2}j_{1}j_{2}}$ and $\delta^{\bot}_{i_{1}i_{2}j_{1}j_{2}}$ for all $(x_{i_1},x_{i_2}) \in {\bf E}(X)$ and all $(y_{j_1},y_{j_2}) \in {\bf E}(Y)$: 
The binary variable   $\delta^{\parallel}_{i_{1}i_{2}j_{1}j_{2}}$ is one if $x_{i_1}$ and $x_{i_2}$ are edges of the vertex-models $\phi(y_{j_1}),\phi(y_{j_2})$ respectively, and the binary variable $\delta^{\bot}_{i_{1}i_{2}j_{1}j_{2}}$ is one if $x_{i_2}$ and $x_{i_1}$ are edges of the vertex-models $\phi(y_{j_1}),\phi(y_{j_2})$ respectively.
This conditions are equivalent to $\delta^{\parallel}_{i_{1}i_{2}j_{1}j_{2}} = \alpha_{i_1j_1}\alpha_{i_2j_2}$ and $\delta^{\bot}_{i_{1}i_{2}j_{1}j_{2}} = \alpha_{i_1j_2}\alpha_{i_2j_1}$.
We can then represent these new variables using linear inequalities as follows: $\quad \forall (x_{i_1},x_{i_2}) \in {\bf E}(X), \forall (y_{j_1},y_{j_2}) \in {\bf E}(Y)$:
\label{deltas_eqs}
    \begin{align}
    \delta^{\parallel}_{i_{1}i_{2}j_{1}j_{2}} &= \alpha_{i_1j_1}\alpha_{i_2j_2} \Leftrightarrow 
    \begin{cases}
        \delta^{\parallel}_{i_{1}i_{2}j_{1}j_{2}} & \leq \alpha_{i_1j_1} \\
        \delta^{\parallel}_{i_{1}i_{2}j_{1}j_{2}} & \leq \alpha_{i_2j_2} \\
        \delta^{\parallel}_{i_{1}i_{2}j_{1}j_{2}} & \geq \alpha_{i_1j_1} + \alpha_{i_2j_2} - 1
    \end{cases} \nonumber
    \end{align}
\noindent and equivalently for $\delta^{\bot}_{i_{1}i_{2}j_{1}j_{2}}$.
Both variables cannot be one for a single combination of $(i_{1}i_{2}j_{1}j_{2})$ simultaneously. This leads to the following constraint.
\begin{equation}
    \delta^{\parallel}_{i_{1}i_{2}j_{1}j_{2}} + \delta^{\bot}_{i_{1}i_{2}j_{1}j_{2}} \leq 1 \quad \forall (x_{i_1},x_{i_2}) \in {\bf E}(X), \forall (y_{j_1},y_{j_2}) \in {\bf E}(Y).
    \label{deltassym_ip}
\end{equation}

\subsection{Complete IP model} 
The feasible region of the IP formulation is defined by:
\begin{equation} \label{feasible_region}
    F = \Big \{(\boldsymbol{\alpha},\boldsymbol{\gamma},\boldsymbol{\delta}^{\parallel},\boldsymbol{\delta}^{\bot}) | (\boldsymbol{\alpha},\boldsymbol{\gamma},\boldsymbol{\delta}^{\parallel},\boldsymbol{\delta}^{\bot}) \in (\eqref{well_defined} \cap \cdots \cap \eqref{deltassym_ip} ) \Big\}.
\end{equation}
%
%
 %
%
A constant objective function can be set for this problem such that any solution that lies within the feasible region defined in Eq.~\eqref{feasible_region} optimizes it.

\noindent\emph{Embedding size.} 
Another choice is given by the {embedding size}:   
Given the limitations on the available quantum annealing hardware in the size of available qubits, a desired property of an embedding is to have a small \textit{qubit footprint}.
The objective function in this case is encoded as 
\begin{equation}
    \min \sum_{i: x_i \in {\bf V}(X)} \sum_{y_j \in {\bf V}(Y)}  \alpha_{ij} \quad 
     \text{ s.t. } (\boldsymbol{\alpha},\boldsymbol{\gamma},\boldsymbol{\delta}^{\parallel},\boldsymbol{\delta}^{\bot}) \in F.
\label{eq:min_size}
\end{equation}

Other objective functions such as fiber size minimization, minimal fiber size dispersion, small difference between a variable degree and fiber size, and available edges in the embedding are also IP representable and can be implemented within this framework.

\section{Decomposition approach}
\label{sec:const_gen}
Implementing all the constraints at once in the IP formulation leads to a model which is often intractable in practice.
The fiber conditions require many constraints to be enforced, and only a small fraction of these are active in optimal solutions.
We investigate the application of a decomposition approach which iterates between a qubit assignment master problem and fiber condition checking subproblems.
The strategy adds strengthened `no-good' constraints (i.e., cuts) to the master problem when they are found to be violated.
Such an approach bears resemblance to decomposition techniques used for scheduling and routing problems, such as classical and logic-based Benders decomposition and branch-and-check \cite{geoffrion1972generalized,hooker2003logic,roshanaei2019branch}.

\subsection{Master Problem}

In the master problem, we relax the fiber conditions, permitting a node in the logical graph to be mapped in multiple parts of the target graph without being connected.
For our master problem, we introduce a new binary decision variable, $z_{e_xe_y} \quad \forall e_x \in {\bf E}(X), \forall e_y \in {\bf E}(Y)$, to track the embedding of problem edges in the target graph edges.
The variable takes on a value of 1 if if edge $e_x = (x_{i_1},x_{i_2}) : e_x \in {\bf E}(X)$ is mapped through the embedding in edge $e_y = (y_{j_1},y_{j_2}) : e_y \in {\bf E}(Y)$, and 0 otherwise.
For modeling purposes, we also denote $e_{x,1} = x_{i_1}, e_{x,2} = x_{i_2}, e_{y,1} = y_{j_1},$ and $e_{y,2} = y_{j_2}$. 
This master problem formulation includes previously expressed mapping constraints, Eq.~\eqref{well_defined}, and size constraints in Eq.~\eqref{size_ip}, in addition to constraints~\eqref{edges_assign} through~\eqref{aggregated} as follows:

\noindent\emph{Assignment of edges.}
Each edge in the source graph has to be assigned to an edge in the target graph.
\begin{equation}
    \sum_{e_x \in {\bf E}(X)} z_{e_xe_y} = 1 \quad \forall e_y \in {\bf E}(Y).
    \label{edges_assign}
\end{equation}

\noindent\emph{Linking constraints.}
To link the assigned qubit values to the $z_{e_xe_y}$ variables, we use the following set of constraints $\forall e_x \in {\bf E}(X), \forall e_y \in {\bf E}(Y)$:
\begin{equation}\label{goals}
    z_{e_xe_y} \leq \alpha_{e_{x,1} e_{x,2}} \quad z_{e_xe_y} \leq \alpha_{e_{y,1} e_{y,2}}.
\end{equation}
Together, these constraints ensure that a problem edge can only be assigned to an edge in the target graph if the pair of nodes involved in that edge take on the required values, which are aggregated in the following constraint
\begin{equation}
    2 \cdot z_{e_xe_y} \leq \alpha_{e_{x,1} e_{x,2}} + \alpha_{e_{y,1} e_{y,2}} \quad \forall e_x \in {\bf E}(X), \forall e_y \in {\bf E}(Y). \label{aggregated}
\end{equation}

\noindent\emph{Subproblem relaxation.}
Although the constraints above already represent the assignment problem to be modeled in the master problem, we can include a relaxation of the subproblem to help guide to master problem towards feasible solutions.
This requires the addition of another set of binary variables, $w_{j}$ that track whether vertex model $\phi(y_j)$ has a size greater than one.
Then, $\forall y_j \in {\bf V}(Y)$:
\begin{subequations}\label{subprob_relax}
\begin{align}
    \sum_{i: x_i \in {\bf V}(X)} \alpha_{ij} - n  \cdot w_j & \leq 1,& \\
    n(1 - \alpha_{ij}) + \sum_{\ell: (x_i,x_\ell) \in {\bf E}(X)} \alpha_{\ell j} + \sum_{\ell: (x_\ell,x_i) \in {\bf E}(X)} \alpha_{\ell j} & \geq w_j
    \quad \forall x_i \in {\bf V}(X).&
\end{align}
\end{subequations} 
This constraints ensure that the variable $w_j$ is one if the node $y_j$ is mapped to more than one node $x_i$.
Eqs. \eqref{edges_assign}-\eqref{subprob_relax}, together with the cuts generated by the subproblems, define the master problem.

\subsection{Subproblems}
The subproblem validates if there exist vertices in the embedding belonging to the same vertex model $\phi(y_j)$ which are not connected in the target graph.
If this is the case, it returns a constraint that either: i) encourages connectivity in future iterations, or ii) removes occurrences of the disconnected vertex models from the graph. 
For each vertex model with more than one vertex on the embedding, it checks at each vertex on the target graph that belongs to that vertex model.
If that vertex does not contain an edge that connects it to another vertex of that vertex model, then the checking procedure returns disconnected.

\subsection{Cuts}\label{sec:cuts}
If a particular vertex model is found to be disconnected in the solution, we add a constraint to remove the current solution and prevent future solutions from having the same disconnectivity. 
Let the set of disconnected vertices in the source graph be denoted as $\hat{j}: y_{\hat{j}} \in \hat{Y}$.
Let the set of vertices in the target graph that belong to this vertex model, $y_{\hat{j}}$, in the current incumbent solution, be represented as the vertex model $\phi(y_{\hat{j}}) \subseteq X$. Let the set of vertices that are adjacent to any vertex in $\phi(y_{\hat{j}})$, but are not assigned value $y_{\hat{j}}$, be denoted $\phi'(y_{\hat{j}})$.
The constraint generated in the current iteration for disconnected qubit $y_{\hat{j}}$ is then given by:

\begin{equation} \label{cuts}
    \left(|\phi(y_{\hat{j}})| - \sum_{i:x_i \in \phi(y_{\hat{j}})} \alpha_{i\hat{j}}\right) + \sum_{i:x_i \in \phi'(y_{\hat{j}})} \alpha_{i\hat{j}} \geq 1.  
\end{equation}
This removes the current infeasible solution from the search space and requires the master problem to: i) include at least one fewer vertex with this vertex model (bracketed term), or ii) include at least one more vertex with this vertex model, among the set of vertices that could improve connectivity (non-bracketed term).

Notice that we reformulated the pullback condition from the Eq.~\eqref{pullback_ip} in terms of $\boldsymbol{\delta}^{\parallel}$ and $\boldsymbol{\delta}^{\bot}$ into the variables $z_{e_xe_y}$ and its corresponding constraints, while the fiber condition is relaxed with the subproblem and cut generation procedure.

Following the intuition in~\citep{cai2014practical}, where the heuristic method tries to obtain embeddings with a small qubit footprint, the default objective function implemented in the master problem is to minimize size, as in Eq.~\eqref{eq:min_size}.
This objective leads the master problem to return compact assignments of variables.
In the case that the feasibility objective is considered within this approach, the optimization procedure is stopped when the first feasible solution is found.

\section{Results}
\label{sec:results}
The model in Section~\ref{sec:IP-constraints} was implemented using the Python-package Pyomo \citep{hart2017pyomo}, which interfaces with several solvers, including open-source and commercial solvers.
The decomposition approach, presented in Section~\ref{sec:const_gen}, is implemented in C++ and uses the CPLEX 12.9 solver~\citep{cplex}. Our  approaches are compared with the D-Wave default heurestic \texttt{minorminer}, introduced in~\cite{cai2014practical} (available at~{\sf github.com/dwavesystems/minorminer}). 
Unless otherwise stated, the monolithic IP method assumes a value of maximum fiber size $k=3$, which is justified for the structured random graphs given their construction.
This provides the monolithic method with an advantage with respect to the decomposition method given that the infeasibility proofs are contingent on the value of $k$.
The results below were obtained using a laptop running Ubuntu 18.04 with an Intel Core i7-6820HQ CMU @ 2.7GHz with 8 threads and 16 GB of RAM.

\subsection{Illustrative Example} \label{sec:ill_example}
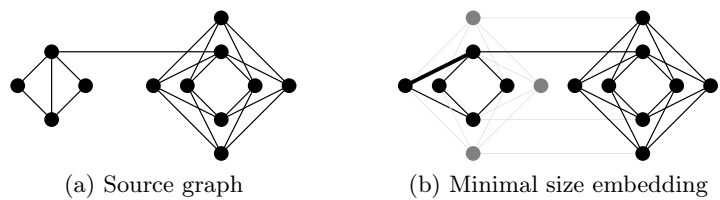
\begin{figure}[hbt!]
    \centering
    \subfloat[Source graph]{\resizebox{!}{2cm}{\begin{tikzpicture}
	\begin{pgfonlayer}{nodelayer}
		\node [style=logical_qubit] (1) at (0, -1) {};
		\node [style=logical_qubit] (4) at (1, 0) {};
		\node [style=logical_qubit] (5) at (-1, 0) {};
		\node [style=logical_qubit] (7) at (0, 1) {};
		\node [style=logical_qubit] (8) at (5, -2) {};
		\node [style=logical_qubit] (9) at (5, -1) {};
		\node [style=logical_qubit] (10) at (5, 2) {};
		\node [style=logical_qubit] (11) at (7, 0) {};
		\node [style=logical_qubit] (12) at (6, 0) {};
		\node [style=logical_qubit] (13) at (4, 0) {};
		\node [style=logical_qubit] (14) at (3, 0) {};
		\node [style=logical_qubit] (15) at (5, 1) {};
	\end{pgfonlayer}
	\begin{pgfonlayer}{edgelayer}
		\draw [style=simple] (7) to (1);
		\draw [style=simple] (15) to (13);
		\draw [style=simple] (15) to (14);
		\draw [style=simple] (15) to (11);
		\draw [style=simple] (9) to (12);
		\draw [style=simple] (8) to (12);
		\draw [style=simple] (9) to (13);
		\draw [style=simple] (13) to (10);
		\draw [style=simple] (13) to (8);
		\draw [style=simple] (8) to (14);
		\draw [style=simple] (14) to (10);
		\draw [style=simple] (14) to (9);
		\draw [style=simple] (8) to (11);
		\draw [style=simple] (9) to (11);
		\draw [style=simple] (11) to (10);
		\draw [style=simple] (1) to (4);
		\draw [style=simple] (7) to (4);
		\draw [style=simple] (7) to (5);
		\draw [style=simple] (15) to (12);
		\draw [style=simple] (7) to (15);
		\draw [style=simple] (1) to (5);
		\draw [style=simple] (12) to (10);
	\end{pgfonlayer}
\end{tikzpicture}}}~~~~~~~~~~~~\subfloat[Minimal size embedding]{\resizebox{!}{2cm}{\begin{tikzpicture}
	\begin{pgfonlayer}{nodelayer}
		\node [style=unused_qubit] (0) at (0, -2) {};
		\node [style=logical_qubit] (1) at (0, -1) {};
		\node [style=unused_qubit] (2) at (0, 2) {};
		\node [style=unused_qubit] (3) at (2, 0) {};
		\node [style=logical_qubit] (4) at (1, 0) {};
		\node [style=logical_qubit] (5) at (-1, 0) {};
		\node [style=logical_qubit] (6) at (-2, 0) {};
		\node [style=logical_qubit] (7) at (0, 1) {};
		\node [style=logical_qubit] (8) at (5, -2) {};
		\node [style=logical_qubit] (9) at (5, -1) {};
		\node [style=logical_qubit] (10) at (5, 2) {};
		\node [style=logical_qubit] (11) at (7, 0) {};
		\node [style=logical_qubit] (12) at (6, 0) {};
		\node [style=logical_qubit] (13) at (4, 0) {};
		\node [style=logical_qubit] (14) at (3, 0) {};
		\node [style=logical_qubit] (15) at (5, 1) {};
	\end{pgfonlayer}
	\begin{pgfonlayer}{edgelayer}
		\draw [style=simple] (15) to (13);
		\draw [style=simple] (15) to (14);
		\draw [style=simple] (15) to (11);
		\draw [style=unused] (0) to (8);
		\draw [style=unused] (1) to (9);
		\draw [style=simple] (9) to (12);
		\draw [style=simple] (8) to (12);
		\draw [style=simple] (9) to (13);
		\draw [style=simple] (13) to (10);
		\draw [style=simple] (13) to (8);
		\draw [style=simple] (8) to (14);
		\draw [style=simple] (14) to (10);
		\draw [style=simple] (14) to (9);
		\draw [style=simple] (8) to (11);
		\draw [style=simple] (9) to (11);
		\draw [style=simple] (11) to (10);
		\draw [style=simple] (6) to (1);
		\draw [style=unused] (3) to (2);
		\draw [style=simple] (1) to (4);
		\draw [style=unused] (0) to (4);
		\draw [style=unused] (4) to (2);
		\draw [style=unused] (5) to (2);
		\draw [style=unused] (5) to (0);
		\draw [style=unused] (6) to (2);
		\draw [style=unused] (0) to (3);
		\draw [style=unused] (1) to (3);
		\draw [style=simple] (7) to (4);
		\draw [style=simple] (7) to (5);
		\draw [style=embedding] (7) to (6);
		\draw [style=unused] (7) to (3);
		\draw [style=simple] (15) to (12);
		\draw [style=unused] (0) to (6);
		\draw [style=simple] (7) to (15);
		\draw [style=unused] (2) to (10);
		\draw [style=simple] (1) to (5);
		\draw [style=simple] (12) to (10);
	\end{pgfonlayer}
\end{tikzpicture}}}
\caption{{\small Source graph of illustrative example~\citep{dridi2018novel} and its minimal size embedding in $\mathcal{C}_{4,1,2}$. Grey nodes and edges represent unused nodes and edges in embedding, but present at the target graph. Bold edges represent edges in chains
.}}
\label{fig:illustrative_example}
\end{figure}

This example is taken from~\citep{dridi2018novel}, where a $K_{4,4}$ bipartite graph is connected through a single edge to a structured 4 nodes graph and is embedded in a $\mathcal{C}_{4,1,2}$ chimera graph as seen in Figure~\ref{fig:illustrative_example}.
This embedding is challenging for heuristic methods that search vertex models outside of the blocks~\citep{dridi2018novel}.
The embedding with the minimal size is given when one of the nodes in the 4-node block is embedded in a chain of length 2, resulting in an embedding of length 13.
The heuristic implemented in \texttt{minorminer} fails ~50\% of the times tried (1000), in the sense that it is not able to find a valid embedding in half of the experiments. 
We consider solving this problem using the CAG approach proposed in~\citep{dridi2018novel}, by computing the Groebner basis of the polynomial ideal.
When using the software Maple 2017~\citep{maple}, which includes Faug\`{e}re's algorithm~\citep{Faugere199961,Faugere:2002:NEA:780506.780516}, the Groebner basis computation is unable to find a solution after 5 hours of computation before it runs out of memory.
We apply our IP approach and include the open-source solvers GLPK 4.61~\citep{glpk} and CBC 2.9.6~\citep{cbc}, and the commercial solvers Gurobi 8.1~\citep{gurobi} and CPLEX 12.9~\citep{cplex}.
Here we set a time limit of one minute per each run. 

The open-source solvers fail to provide feasible solutions within the time limit when there is a constant objective function.
CBC can find a solution when we consider the embedding size minimization as an objective, showing how the including an objective function can be beneficial for the IP solvers.
In that case, the solver is unable to guarantee the optimality solution, although it finds the optimal solution, and it provides an optimality gap of 8.3\%.
The commercial solvers, on the other hand, can provide both feasible and optimal solutions in under a minute of computation.
In particular, Gurobi takes 1.3 seconds to find a feasible solution and 31.2 seconds to find and prove the optimality of the solution while CPLEX takes 3.5 seconds and 9.4 seconds in the same tasks, respectively.
As expected, the time required to provide a feasible solution is less than that to give optimality guarantees.
Based on the better results obtained using CPLEX compared to Gurobi, we only present the results using this solver in the remaining of the manuscript.
Apart from it, the decomposition approach can provide feasible solutions more efficiently, and with higher quality compared to the other approaches.
In this case, it only took 0.4 seconds for the decomposition approach to provide a feasible solution which was nearly optimal, with a 7\% optimality gap, and to find a provable optimal solution it required 46.5 seconds.
These results suggest that the usage of commercial solvers is required for solving these challenging IP problems, and therefore the results presented in the remaining of this manuscript correspond to these solvers.

\subsection{Random Graphs}

\noindent\emph{1. Random structured graphs.}
Here we generalize the example above. We consider the bipartite graph $K_{4,4}(p_{inter})$ parametrized by $p_{inter}$, which is the probability of the existence of edges between the two partitions. We randomly choose
$\zeta$ edges, which we {contract} into nodes (each edge into a single node). This graph is then connected (attached) to a complete $K_{4,4}$ bipartite graph by 4 edges chosen with a probability $p_{intra}$.
By construction, the resulting graph is a subgraph of  $\mathcal{C}_{4,1,2}$, and its size is $m + \zeta$. It is the smallest minor of the corresponding graph without contraction. The example of Section~\ref{sec:ill_example} is obtained with $\zeta=1$ (and $m=12$).

\begin{figure}[hbt!]
    \centering
    \subfloat[Embedding time]{
        \includegraphics[height=0.5\textwidth,valign=t]{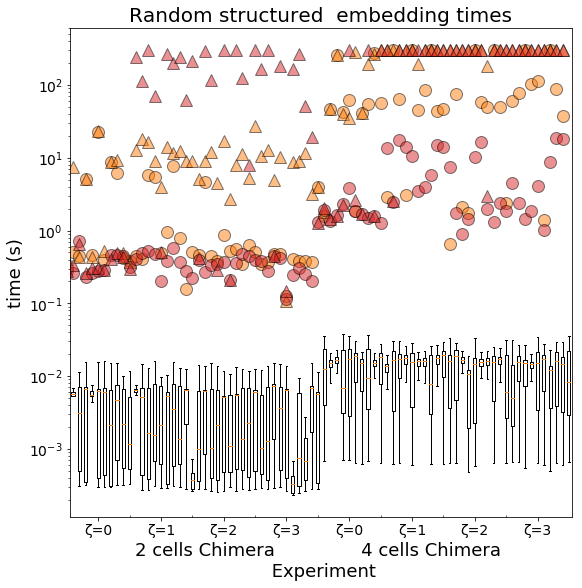}
        \label{fig:structured_time}
    }%
    \subfloat[Embedding size]{
        \includegraphics[height=0.5\textwidth,valign=t]{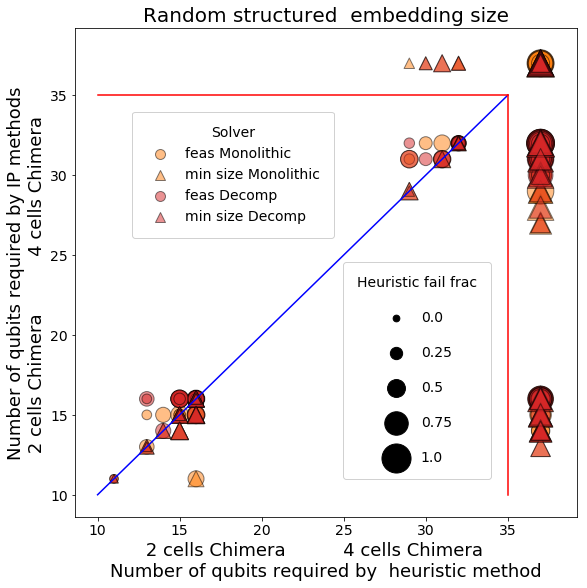}
        \label{fig:structured_size}
    }
    \caption{{\small Embedding time and size comparison for different embedding methods for structured random graphs in $\mathcal{C}_{4,1,2}$ and $\mathcal{C}_{4,2,2}$ with respect to median behavior of \texttt{minorminer}. Values beyond the red lines represent embeddings where the heuristic median performance (right) or the IP methods (top) failed to return an embedding.}}
    \label{fig:structured}
\end{figure}

Fixing $p_{inter}=p_{intra}=0.5$, for each value of of contracted edges $\zeta \in \{0,\cdots,3\}$, we generated 10 random graphs.
These random graphs were embedded in a $\mathcal{C}_{4,1,2}$ graph with a time limit of 300 seconds. 
Figure~\ref{fig:structured_time} gives the runtimes for the monolithic IP  and the decomposition methods solved using CPLEX. 
This figure also shows the boxplots for the 1000 runs of \texttt{minorminer}.
For this case, given the way the random structured graphs are constructed, we see that the longest fiber will be at most of size 3, which we encode for the monolithic IP approach using the parameter $k=3$.
Notice that this observation biases the results in favor of the monolithic IP approach with respect to the decomposition approach.
For finding a feasible solution, the decomposition approach is more efficient than the monolithic IP approach. When $\zeta=0$, where finding a feasible embedding is practically finding the minimally sized embedding, there is no difference in time performance between the cases of embedding size minimization and finding a feasible solution.
For $\zeta>0$, the embedding size minimization becomes more expensive, in particular for the decomposition approach.
The monolithic IP and the decomposition approaches were able to find smaller or equally sized embeddings for 33 and 30 cases out of the 40 experiments, respectively. When the objective function is
the size minimization, this number increased for all instances in all cases and is strictly better in 22 cases for the monolithic IP and 21 cases for the decomposition approach.
Notice that the monolithic IP approach was able to find an embedding for one instance which was smaller than any of the 1000 runs of the heuristic method.

Larger instances of random structures graphs can be generated by combining two graphs like the ones described above, and include the edges appearing in a $\mathcal{C}_{4,2,2}$ graph between the cells with probability $p_{intra}$.
As before, we generated 10 random instances with values of $\zeta$ ranging from 0 to 3.
The time performance of the different methods is shown in Figure~\ref{fig:structured_time}. 
Out of the 40 instances the monolithic IP and the decomposition method are still able to find embeddings as succinct as the median heuristic behavior in 17 and 23 instances when trying to find a feasible solution, and in 20 and 17 instances when trying to minimize the size of the embedding, respectively.
As in the previous case, the monolithic IP approach is able to find smaller embeddings than any of the 1000 runs of the heuristic method, although for this family of instances this happens for two cases.


Figure~\ref{fig:structured_size} shows a comparison of the embedding median sizes obtained by the heuristic method versus the ones obtained for the IP methods.
The size of the makers represents the heuristic failure rate fraction, computed from the 1000 runs of the heuristic method for each instance.
Both the monolithic and the decomposition approaches (different colors) and the feasibility and size minimization objectives (different markers) are represented in this figure.
In total, out of the 80 structured instances, the heuristic failed more than 50\% for 47 instances, more than 80\% for 12 instances, and more than 90\% for 3 instances.
These instances appear on the right size of Figure~\ref{fig:structured_size}.
For those instances, the IP approaches were able to find a feasible solution in less than 20 seconds, and only for 19, 7, and 2 instances, respectively, the optimal solution could not be guaranteed within the time limit.
For high failure rate problems~{($>50\%$ failure rate)} for the heuristic, our methods find a feasible solution in under 20 seconds and prove optimality in more than half of the instances in less than 5 minutes.
Notice that most of the runs corresponding to finding a feasible embedding (circles) are above the diagonal line, indicating a larger embedding size for the IP methods compared to the embeddings, while size minimization runs (triangles) lie on the diagonal or below it. 

\noindent\emph{2. Erd\"os-R\'enyi graphs.}
These graphs are parametrized by the number of nodes $\nu$ and the probability of an edge existing between each pair of nodes $p$.
We consider a set of 10 random instances for each combination of $\nu \in \{5,6,\cdots,16\}$ and $p \in \{0.3,0.5,0.7\}$.
Each of this graphs is embedded in different sizes of Chimera, $\mathcal{C}_{4,1,1},\mathcal{C}_{4,2,1},\mathcal{C}_{4,3,1},$ and $\mathcal{C}_{4,2,2}$, and Pegasus, $\mathcal{P}_{4,1,1,1}$ and $\mathcal{P}_{4,2,2,3}$. We set the 
 time limit to 60 seconds.
In the trivially infeasible case where $\nu > n$ our methods could almost immediately identify the infeasibility, contrary to the \texttt{minorminer} heuristic.
The conclusion is that  the runtime for the monolithic  IP methods increases with the size of the target and source graphs, the density of the source graph given by $p$, and when the objective function is to minimize the embedding sizes.

In this experiment, we considered  2160 instances. In 1100 of them,  the heuristic method could not find any feasible embedding after 1000 runs--
Expectedly, the number of infeasible embeddings significantly drops when using the Pegasus graph.
In 94\% of these cases, at least one of the monolithic  IP methods does not time out, meaning that the methods could prove the infeasibility of the embedding or find a feasible embedding.
This proves that the methods proposed in this work are valid for providing guarantees of embeddability of graph minors in cases where the current heuristics are unable to answer this satifiability question.

We complete our benchmark of random graphs embedding larger problems.
In this case we, consider 5 random instances for each combination of $\nu \in \{10,15,\cdots,35\}$ and $p \in \{ 0.1,0.3,0.5,0.7\}$ embedded into $\mathcal{C}_{4,4,4}$, where the longest fiber size was increased to $k=5$.
We observe that for these instances, the only IP solver that does not run out time is CPLEX implementing either the monolithic IP or the decomposition approach with the feasible solution objective.

\begin{figure}[h!]
    \centering
    \subfloat[Feasibility: Time ]{\includegraphics[width=0.32\textwidth]{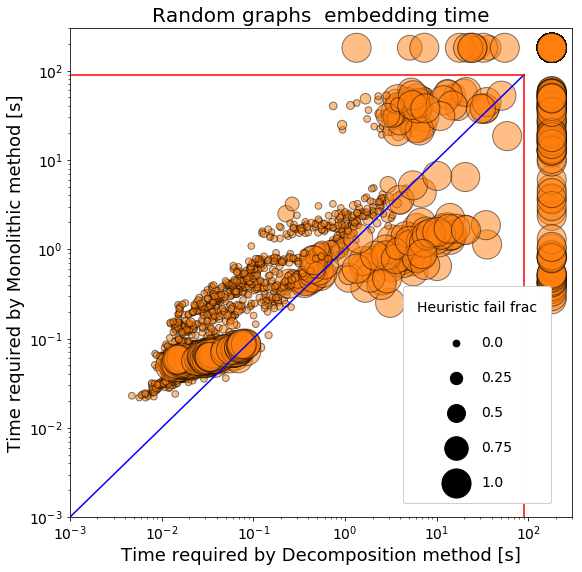}
    \label{fig:small_time_feas_compare}}
    \subfloat[Minimize size: Size]{\includegraphics[width=0.32\textwidth]{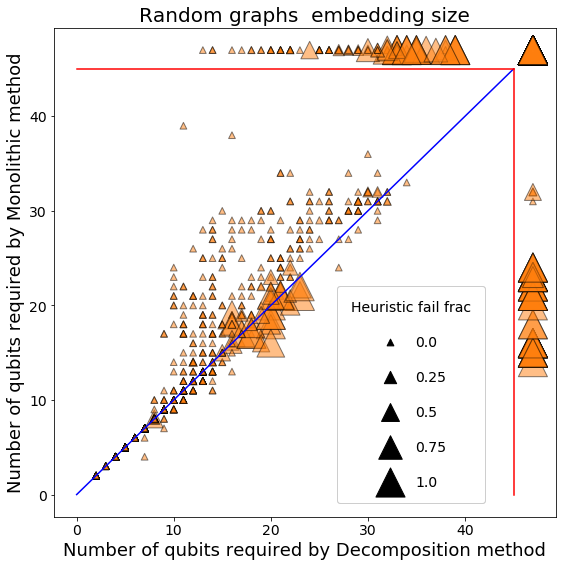}
    \label{fig:small_size_min_compare}}%
    \subfloat[Minimize size: Time]{\includegraphics[width=0.32\textwidth]{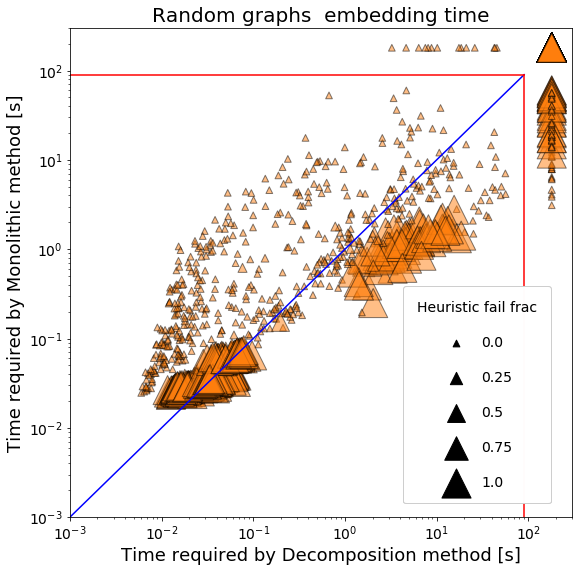}
    \label{fig:small_time_min_compare}}
    \caption{{\small Embedding size and time comparison for Erd\"os-R\'enyi graphs ($\nu \in \{5,6,\cdots,16\}$, $p \in \{0.3,0.5,0.7\}$) given different objectives. Values beyond the red lines represent embedding where the decomposition (right) or the monolithic IP methods (top) failed to return an embedding or went over the time limit. 
    }}
    \label{fig:small_compare}
\end{figure}

Figure~\ref{fig:small_compare}  presents the embedding size and time comparison for the small random graph experiments. 
For this test-case, in 60\% of the instances,
the decomposition approach yielded embeddings with sizes equal or smaller than the median of the ones returned by the heuristic, when looking for a feasible solution, and in 90\% of the instances when minimizing the embedding size.
The monolithic IP approach was more efficient to declare infeasibility in non-trivial cases ($m<n$) than the decomposition approach, Figures \ref{fig:small_time_feas_compare} and \ref{fig:small_size_min_compare}, as the values below the diagonal with large heuristic failure fractions and longer runtimes.
When compared to the minimal size found after the 1000 runs of the heuristic method, the monolithic IP methods are still able to find smaller embeddings for around 5\% of the cases.
The comparison in Figure~\ref{fig:small_compare} highlights that the sizes of the embeddings found by the decomposition approach are in most cases as small or smaller than the monolithic IP approach.
In terms of the computational time, we observe that those instances that were challenging for the heuristic (large markers) are more easily solved by the decomposition approach, especially when minimizing the size of the embedding.
The remaining instances appear to be solved more efficiently using the monolithic IP approach.
The larger and more challenging instances lead to different results.
Out of the 120 instances solved, the decomposition approach behaves better than the monolithic IP approaches, obtaining equally good or better embedding than the median heuristic behavior in 30\% of the cases, compared to around 10\% for the monolithic IP approaches.
The solution requires larger fibers, which affected directly the formulation size of the monolithic case making it more challenging to solve.
Only for one instance, a smaller embedding than any of the observed heuristic solutions is obtained, in this case by the decomposition approach.
\subsection{Applications}
\noindent\emph{1. Gadgets. } 
It has been shown in \citep{dattani2019embedding}, that all the cubic gadgets can be embedded in a single cell of either Chimera of Pegasus graphs, but three of the quartic gadgets required more than a single Chimera cell.
The three gadgets were $K_6-e$, $Double~K_4$, and $K_6$.
%
%
We find more efficient embeddings for several of the gadgets, namely $K_5$ (with 2 and 1 auxiliary variables), $K_6$, and $K_6 - e$ compared to~
\citep{dattani2019embedding}.
The embeddings found can be guaranteed to be the minimal size within a few seconds of computation. 
For the case of the quartic gadgets, all but one (double~$K_4$) could be embedded in a single Chimera cell, in which case our method could provide infeasibility guarantees in less than 10 seconds.
%

\noindent\emph{2. Spanning tree. }
An example of an application is the communication of vehicles/agents with a central control station that can be disrupted in a particular area and can be routed through $\Delta$   agents/vehicles.
Finding the communication routing of the vehicles that minimizes the distance, is equivalent to finding the minimum spanning tree with bounded degree $\Delta$.
\citet{rieffel2019ans} propose three different formulations of this problem that can be embedded in a quantum annealer.
Given the graph defined by the agents/vehicles $S = (V,E)$, the distances among them might change but not the graph itself.
At the same time, the degree of the spanning tree $\Delta$ is fixed by the communication equipment.
We generate instances 80 with graphs $S$ with 4 vertices and between 3 and 5 edges.
When reformulating the problems as QUBOs, we obtain instances where the target graph ranges in size between 19 and 29 nodes, and with 35 to 70 edges.
The resulting QUBOs were embedded using the decomposition approach and compared to the heuristic in \texttt{minorminer}.
We obtain smaller (or equally sized) embeddings  than the median length of the heuristic in 12.5\% (15\%) of the instances in 5 minutes of computational time (compared to 1000 runs of the heuristic).


\noindent\emph{3. Protein folding. }
\citet{Perdomo2012} encode the different configuration of the amino acids in a protein in terms of a QUBO representing the overall energy of the system.
Minimizing this QUBO with respect to the different number of bonds between amino-acids would yield the protein's least-energy configuration.
\citep{Perdomo2012} shows not only the encoding of the problem as a QUBO, but also provides a custom algorithm to embed it in the D-Wave One chip, which has hardware described by a faulty $\mathcal{C}_{4,4,4}$ graph.
The largest instance solved in this paper involved embedding a QUBO of 19 variables in a target graph of 127 qubits.
The resulting embedding involved 81 qubits with the largest vertex model of length 5, being at the time the largest problem embedded and solved in D-Wave's quantum annealers.
We highlight two qualities of our approach: 1) we are not making any assumption about the source or target graphs, allowing us to work with faulty Chimera graphs as targets; and 2) we can exploit the fact of having an existing embedding to initialize our procedures, allowing us to solve our IP problems more efficiently.
Initializing with the embedding provided by \citet{Perdomo2012} while restricting the $k=5$ in the monolithic IP approach, 
we find an embedding of length 77 (4.9\% qubit footprint reduction) within an hour of computation.
Allowing larger fibers, we find an embedding of size 74 (8.6\% qubit footprint reduction) with a fiber of size 6.
These embeddings are not guaranteed to be optimal, but in both cases improve those previously found.

%
%
%

\section{Conclusions}
\label{sec:conclusions}
Integer programming (IP) approaches are proposed to solve the graph minor-embedding problem. Specifically, we develop a monolithic IP derived from the polynomial equations presented in \citet{dridi2018novel}, and a decomposition approach, both of which are capable of identifying infeasible instances and providing bounds on solution quality. These approaches are also agnostic of the source and target graphs. 
Both approaches were implemented and tested using a range of different source graphs with various sizes, densities, and structures. The target graphs used follow the architecture of the chips in D-Wave's current and future quantum annealers.
Although slower overall than the currently-employed heuristic method \cite{cai2014practical}, the proposed methods prove to be a viable solution approach for highly structured source graphs, where the heuristic fails with a higher probability.

The results presented in this paper highlight the more general approaches to minor-embedding using IP.
Another way of obtaining better performance is by reducing the search space by imposing certain limitations to the embedding, e.g. by allowing only certain topologies for the vertex models or by fixing certain embedding characteristics, like maximum fiber size. Initial attempts to include these approximations show a promising decrease in the computation time with an acceptable trade-off in quality. Our formulation and results are a baseline for future methods that can work at application-scale. A promising future direction is to use symmetries and the invariant formulation as previously suggested \cite{dridi2018novel}. Finally, applications in gadget embeddings, spanning tree problems, and protein folding demonstrate the advantages of our approaches.

\subsection*{Acknowledgements}
We thank Prof. Ignacio Grossmann  and Dr. Eleanor Rieffel for the constructive discussions during the preparing of this work.
DB and DV are supported/partially supported by USRA Feynman Quantum Academy, NASA NAMS (NNA16BD14C), AFRL NYSTEC Contract (FA8750-19-3-6101).
DB is also supported by the Center of Advanced Process Decision Making (CAPD) at CMU.

\renewcommand{\bibname}{{\leftline{References}}}
\bibliographystyle{plainnat}
\bibliography{references}

\def\cprime{$'$} \def\cprime{$'$}
\begin{thebibliography}{29}
\providecommand{\natexlab}[1]{#1}
\providecommand{\url}[1]{\texttt{#1}}
\expandafter\ifx\csname urlstyle\endcsname\relax
  \providecommand{\doi}[1]{doi: #1}\else
  \providecommand{\doi}{doi: \begingroup \urlstyle{rm}\Url}\fi

\bibitem[Bian et~al.(2014)Bian, Chudak, Israel, Lackey, Macready, and
  Roy]{bian2014discrete}
Zhengbing Bian, Fabian Chudak, Robert Israel, Brad Lackey, William~G Macready,
  and Aidan Roy.
\newblock Discrete optimization using quantum annealing on sparse {I}sing
  models.
\newblock \emph{Frontiers in Physics}, 2:\penalty0 56, 2014.

\bibitem[Bixby(2012)]{Bixby2012}
Robert~E Bixby.
\newblock {A Brief History of Linear and Mixed-Integer Programming
  Computation}.
\newblock \emph{Documenta Mathematica {\textperiodcentered} Extra}, 2012.

\bibitem[Boothby et~al.(2019)Boothby, Bunyk, Raymond, and Roy]{boothby2019next}
Kelly Boothby, Paul Bunyk, Jack Raymond, and Aidan Roy.
\newblock Next-generation topology of {D}-wave quantum processors.
\newblock Technical report, Technical report, 2019.

\bibitem[Cai et~al.(2014)Cai, Macready, and Roy]{cai2014practical}
Jun Cai, William~G Macready, and Aidan Roy.
\newblock A practical heuristic for finding graph minors.
\newblock \emph{arXiv:1406.2741}, 2014.

\bibitem[Choi(2011)]{Choi2011}
Vicky Choi.
\newblock Minor-embedding in adiabatic quantum computation: {II}.
  minor-universal graph design.
\newblock \emph{Quantum Information Processing}, 10\penalty0 (3):\penalty0
  343--353, 2011.

\bibitem[Cox et~al.(1998)Cox, Little, and O'Shea]{cox}
David~A. Cox, John~B. Little, and Donal O'Shea.
\newblock \emph{Using algebraic geometry}.
\newblock Graduate texts in mathematics. Springer, New York, 1998.
\newblock ISBN 0-387-98487-9.

\bibitem[Cplex(2019)]{cplex}
Cplex.
\newblock 12.9 user’s manual.
\newblock 2019.

\bibitem[Dattani and Chancellor(2019)]{dattani2019embedding}
Nike Dattani and Nick Chancellor.
\newblock Embedding quadratization gadgets on {C}himera and {P}egasus graphs.
\newblock \emph{arXiv:1901.07676}, 2019.

\bibitem[Dridi et~al.(2018)Dridi, Alghassi, and Tayur]{dridi2018novel}
Raouf Dridi, Hedayat Alghassi, and Sridhar Tayur.
\newblock A novel algebraic geometry compiling framework for adiabatic quantum
  computations.
\newblock \emph{arXiv:1810.01440}, 2018.

\bibitem[Dridi et~al.(2019)Dridi, Alghassi, and Tayur]{dridiAGIntro}
Raouf Dridi, Hedayat Alghassi, and Sridhar Tayur.
\newblock Minimizing polynominal functions on quantum computers.
\newblock \emph{{S}cience and {C}ulture, MAY-JUNE}, 2019.

\bibitem[Faug\`{e}re(1999)]{Faugere199961}
Jean-Charles Faug\`{e}re.
\newblock A new efficient algorithm for computing {G}r\"obner bases ({F4}).
\newblock \emph{Journal of Pure and Applied Algebra}, 139\penalty0
  (13):\penalty0 61 -- 88, 1999.

\bibitem[Faug\`{e}re(2002)]{Faugere:2002:NEA:780506.780516}
Jean~Charles Faug\`{e}re.
\newblock A new efficient algorithm for computing {G}r\"obner bases without
  reduction to zero ({F5}).
\newblock In \emph{Proceedings of the 2002 International Symposium on Symbolic
  and Algebraic Computation}, ISSAC '02, pages 75--83, New York, NY, USA, 2002.
  ACM.

\bibitem[Forrest and Lougee-Heimer(2005)]{cbc}
John Forrest and Robin Lougee-Heimer.
\newblock {CBC User Guide}.
\newblock In \emph{Emerging Theory, Methods, and Applications}. 2005.

\bibitem[Geoffrion(1972)]{geoffrion1972generalized}
Arthur~M Geoffrion.
\newblock Generalized benders decomposition.
\newblock \emph{Journal of optimization theory and applications}, 10\penalty0
  (4):\penalty0 237--260, 1972.

\bibitem[Goodrich et~al.(2018)Goodrich, Sullivan, and
  Humble]{goodrich2018optimizing}
Timothy~D Goodrich, Blair~D Sullivan, and Travis~S Humble.
\newblock Optimizing adiabatic quantum program compilation using a
  graph-theoretic framework.
\newblock \emph{Quantum Information Processing}, 17\penalty0 (5):\penalty0 118,
  2018.

\bibitem[Gurobi~Optimization(2019)]{gurobi}
LLC Gurobi~Optimization.
\newblock Gurobi optimizer reference manual, 2019.

\bibitem[Hart et~al.(2017)Hart, Laird, Watson, Woodruff, Hackebeil, Nicholson,
  and Siirola]{hart2017pyomo}
William~E Hart, Carl~D Laird, Jean-Paul Watson, David~L Woodruff, Gabriel~A
  Hackebeil, Bethany~L Nicholson, and John~D Siirola.
\newblock \emph{Pyomo-optimization modeling in {P}ython}, volume~67.
\newblock Springer, 2017.

\bibitem[Hooker and Ottosson(2003)]{hooker2003logic}
John~N Hooker and Greger Ottosson.
\newblock Logic-based {B}enders decomposition.
\newblock \emph{Mathematical Programming}, 96\penalty0 (1):\penalty0 33--60,
  2003.

\bibitem[Kaminsky and Lloyd(2004)]{kaminsky2004scalable}
William~M Kaminsky and Seth Lloyd.
\newblock Scalable architecture for adiabatic quantum computing of {NP}-hard
  problems.
\newblock In \emph{Quantum computing and quantum bits in mesoscopic systems},
  pages 229--236. Springer, 2004.

\bibitem[Maplesoft(2019)]{maple}
Maplesoft.
\newblock Algorithms for {G}roebner basis, {M}aple 2017.
\newblock 2019.

\bibitem[Neven et~al.(2009)Neven, Denchev, Drew-Brook, Zhang, Macready, and
  Rose]{neven2009nips}
Harmut Neven, Vasil~S Denchev, Marshall Drew-Brook, Jiayong Zhang, William~G
  Macready, and Geordie Rose.
\newblock {NIPS} 2009 demonstration: Binary classification using hardware
  implementation of quantum annealing.
\newblock 2009.

\bibitem[Okada et~al.(2019)Okada, Ohzeki, Terabe, and
  Taguchi]{okada2019improving}
Shuntaro Okada, Masayuki Ohzeki, Masayoshi Terabe, and Shinichiro Taguchi.
\newblock Improving solutions by embedding larger subproblems in a {D}-wave
  quantum annealer.
\newblock \emph{Scientific reports}, 9\penalty0 (1):\penalty0 2098, 2019.

\bibitem[Oki(2012)]{glpk}
Eiji Oki.
\newblock {GLPK (GNU Linear Programming Kit)}.
\newblock In \emph{Linear Programming and Algorithms for Communication
  Networks}. 2012.

\bibitem[Perdomo-Ortiz et~al.(2012)Perdomo-Ortiz, Dickson, Drew-Brook, Rose,
  and Aspuru-Guzik]{Perdomo2012}
Alejandro Perdomo-Ortiz, Neil Dickson, Marshall Drew-Brook, Geordie Rose, and
  Alan Aspuru-Guzik.
\newblock {Finding low-energy conformations of lattice protein models by
  quantum annealing}.
\newblock \emph{Scientific Reports}, 2012.

\bibitem[Rieffel et~al.(2019)]{rieffel2019ans}
Eleanor~G Rieffel et~al.
\newblock {From Ans{ä}tze to Z-gates: a NASA View of Quantum Computing}.
\newblock \emph{arXiv:1905.02860}, 2019.

\bibitem[Roshanaei et~al.(2019)Roshanaei, Booth, Aleman, Urbach, and
  Beck]{roshanaei2019branch}
Vahid Roshanaei, Kyle E.~C. Booth, Dionne~M Aleman, David~R Urbach, and
  J~Christopher Beck.
\newblock Branch-and-check methods for multi-level operating room planning and
  scheduling.
\newblock \emph{International Journal of Production Economics}, 2019.

\bibitem[Serra et~al.(2019)Serra, Huang, Raghunathan, and Bergman]{Serra2019}
Thiago Serra, Teng Huang, Arvind Raghunathan, and David Bergman.
\newblock {Template-based Minor Embedding for Adiabatic Quantum Optimization}.
\newblock \emph{arXiv:1910.02179}, 2019.

\bibitem[Teramoto et~al.(2018)Teramoto, Nakamura, Takigawa, Minato, Yamaoka,
  and Komatsuzaki]{teramoto2018graph}
Hiroshi Teramoto, Atsuyoshi Nakamura, Ichigaku Takigawa, Shin-Ichi Minato,
  Masanao Yamaoka, and Tamiki Komatsuzaki.
\newblock Graph minors from simulated annealing for annealing machines with
  sparse connectivity.
\newblock In \emph{Theory and Practice of Natural Computing: 7th International
  Conference, TPNC 2018, Dublin, Ireland, December 12--14, 2018, Proceedings},
  volume 11324, page 111. Springer, 2018.

\bibitem[Yang and Dinneen(2016)]{yang2016graph}
Zongcheng Yang and Michael~J Dinneen.
\newblock Graph minor embeddings for {D}-wave computer architecture.
\newblock Technical report, Department of Computer Science, The University of
  Auckland, New Zealand, 2016.

\end{thebibliography}

\newpage
\appendix
\subsection*{Crash course on Groebner bases}
For completeness, we include the following introductory material adopted from~\citep{dridiAGIntro, dridi2018novel}. First, the notations: 
We write $\mathbb Q[x_0, \ldots, x_{n-1}]$
for the ring of polynomials in $x_0, \ldots, x_{n-1}$ with rational coefficients. Let    
$\mathcal S$ be 
 a  set of polynomials $f\in \mathbb Q[x_0, \ldots, x_{n-1}]$. Let  $\mathcal V(S)$ denotes the  algebraic variety defined by 
the polynomials $f\in S$, that is, the set of common zeros of the equations $f=0, \, f\in \mathcal S$. The system $\mathcal S$ 
generates an {\it ideal} $\mathcal I $ by  taking all linear combinations over $\mathbb Q[x_0, \ldots, x_{n-1}]$ of all polynomials in  $\mathcal S$; we have $\mathcal V(\mathcal S)=\mathcal V(\mathcal I).$ The ideal $\mathcal I$ reveals the hidden polynomials that are the consequence of the generating  polynomials in $\mathcal S$. For instance, if one of the hidden polynomials is the constant polynomial 1 (i.e., $1\in \mathcal I$), then the system $\mathcal S$ is inconsistent (because $1\neq 0$).  To be precise, the set of all hidden polynomials is given
by the so-called {\it radical ideal} $\sqrt{\mathcal I}$, which is defined by \mbox{$\sqrt{\mathcal I} = \{g \in\mathbb Q[x_1, \ldots, x_n] |\,  \exists r\in \mathbb N: \, g^r\in \mathcal I \}$}. We have 
   $\mathcal I(\mathcal V(\mathcal I))=\sqrt{\mathcal I}$. Of course, the radical ideal $\sqrt{\mathcal I}$ is infinite. Luckily, thanks to a prominent technical result (i.e., {\it Dickson's lemma}), it has a finite generating set i.e., a  
 {\it  Groebner basis} $\mathcal B$, which one might take to be a triangularization of the ideal $\sqrt{\mathcal I}$.  The computation of   Groebner bases generalizes Gaussian elimination in linear systems. We also continue to have
  $\mathcal V(\mathcal S)=\mathcal V(\mathcal I)=\mathcal V(\sqrt{\mathcal I})=\mathcal V(\mathcal B)$. 
 Instead of giving the technical definition of what a Groebner basis is (which can be found in~\citep{cox} and in many other textbooks)   let us give an example (for simplicity, we use the term  ``Groebner bases" to refer to  {\it reduced} Groebner bases, which is,  technically what we are working with):
\begin{example}
 Consider the system by 
 $$
     \mathcal S = \{x^2+y^2+z^2-4, x^2+2y^2-5, xz-1\}. 
 $$
 We want to solve $\mathcal S$. One way to do so is to compute a Groebner basis for~$\mathcal S$.
 In Figure~\ref{code1}, the output of cell number 4 gives a Groebner basis of $\mathcal S$. We can see that the initial system has been triangulized:  The last equation contains only the variable $z$, whilst the second has an additional variable, and so on. The variable $z$ is said to be eliminated with respect to the rest of the variables. When computing the Groebner basis,  the underlying algorithm (Buchberger's algorithm)  uses the ordering $x>y>z$ (called lexicographical ordering) for the computing of two internal calculations: cross-multiplications and Euclidean divisions.    The program tries to isolate $z$ first, then $z$ and $y$, and finally $x, y$, and $z$~(all variables). Different orderings yield different Groebner bases.

     \begin{figure}[h]
        \centering
            {{\includegraphics[height=8cm, width=16cm]{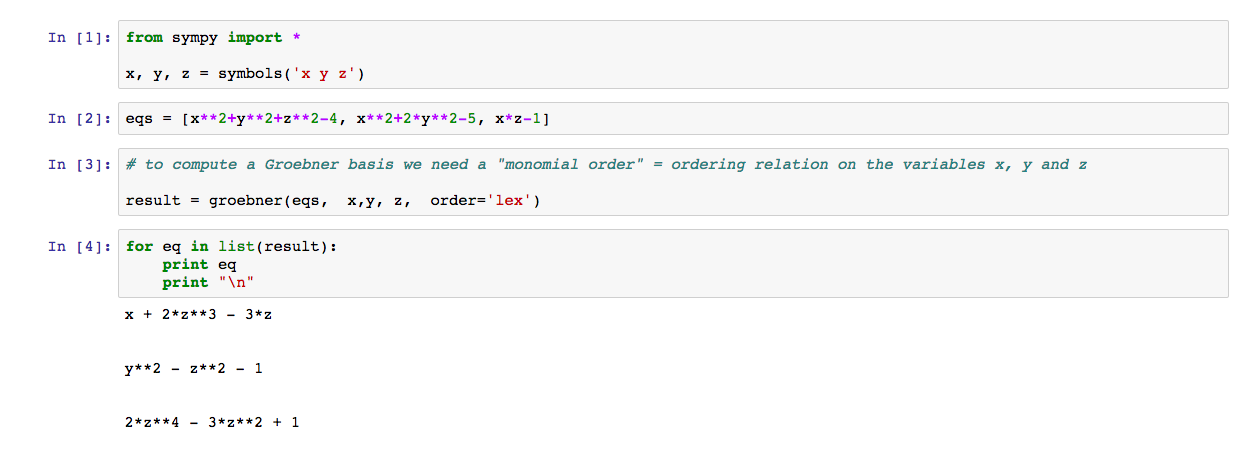} }}
            \caption{{\small Jupyter notebook for computing Groebner bases using   Python package {\sf sympy}. More efficient algorithms exist (e.g.,~\citep{Faugere:2002:NEA:780506.780516, Faugere199961}).  }}
           \label{code1}
      \end{figure}
      
\end{example}

The mathematical power of Groebner bases doesn't stop at solving systems of algebraic equations. The applicability of Groebner bases goes well beyond this:   it gives necessary and sufficient conditions for the existence of solutions.  For instance, if the ideal represents a system of algebraic equations and these equations are (algebraically) dependent on certain parameters, then the  intersection (\ref{intersectionB})
gives {\it all} necessary and sufficient conditions for the existence of solutions.   The following embedding example makes this more concrete. The example also illustrates the original equational approach \cite{dridi2018novel}.

\begin{example}
    Consider the two graphs in Figure~\ref{simpleExple}. We would like to determine all embeddings $\pi:X\rightarrow Y$.   In this case, well-definition equations (I) are 
     given by
      \begin{eqnarray}
     && \alpha_{{1,1}}\alpha_{{1,2}}, \, \alpha_{{1,1}}\alpha_{{1,3}}, \, 
\alpha_{{1,2}}\alpha_{{1,3}},
\\
&&
\alpha_{{2,1}}\alpha_{{2,2}},\, \alpha_{{2,1}}\alpha_{{2,3}},\, \alpha_{{2,2}}\alpha_{{2,3}},
\\
&&
\alpha_{{3,1}}\alpha_{{3,2}},\, \alpha_{{3,1}}\alpha_{{3,3}},\, \alpha_{{3,2}}\alpha_{{3,3}},
\\
&&
\alpha_
{{4,1}}\alpha_{{4,2}},\, 
\alpha_{{4,1}}\alpha_{{4,3}},\, \alpha_{{4,2}}
\alpha_{{4,3}},
\\
&&
\alpha_{{5,1}}\alpha_{{5,2}},\, \alpha_{{5,1}}\alpha_{{5,3}},\, \alpha_{{5,2}}\alpha_{{5,3}}, 
    \end{eqnarray}
and 
\begin{eqnarray}\nonumber
&&
\alpha_{{1,1}}+\alpha_{{1,2}}+\alpha_{{1,3}}-\beta_{{1}},
\quad 
\alpha_{{2,1}}+\alpha_{{2,2}}+\alpha_{{2,3}}-\beta_{{2}},
\quad 
\alpha_{{3,1}
}+\alpha_{{3,2}}+\alpha_{{3,3}}-\beta_{{3}},
\\\nonumber
&&
\alpha_{{4,1}}+\alpha_{{4,
2}}+\alpha_{{4,3}}-\beta_{{4}}, 
\quad
\alpha_{{5,1}}+\alpha_{{5,2}}+\alpha_{{
5,3}}-\beta_{{5}}. 
\end{eqnarray}
  \begin{figure}[h] %
    \centering
    \subfloat 
    {{\includegraphics[width=4cm]{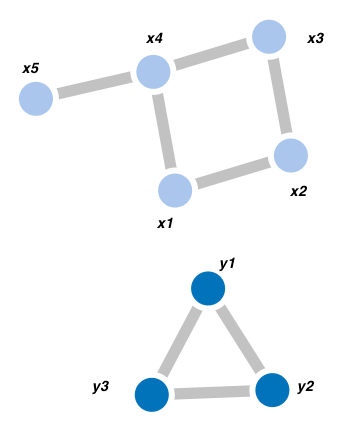} }}%
    \caption[]{{The set of  {\it all} fiber bundles $\pi:X\rightarrow Y$
    defines an algebraic variety. This variety is given by the Groebner basis (\ref{GBSimpleExple}).
    }}%
    \label{simpleExple}%
\end{figure}

The {pullback condition} reads
{\small 
\begin{eqnarray}\nonumber
    &&-1+\alpha_{{4,1}}\alpha_{{5,2}}+\alpha_{{3,1}}\alpha_{{4,2}}+
\alpha_{{1,1}}\alpha_{{2,2}}+\alpha_{{3,2}}\alpha_{{4,1}}+\alpha_{{1,2
}}\alpha_{{2,1}}+\alpha_{{1,2}}\alpha_{{4,1}}+\alpha_{{2,2}}\alpha_{{3
,1}}+\alpha_{{1,1}}\alpha_{{4,2}}+\alpha_{{2,1}}\alpha_{{3,2}}+\alpha_
{{4,2}}\alpha_{{5,1}},
\\\nonumber
&&
-1 +\alpha_{{3,3}}\alpha_{{4,1}}+\alpha_{{1,3}}
\alpha_{{2,1}}+\alpha_{{2,3}}\alpha_{{3,1}}+\alpha_{{4,1}}\alpha_{{5
,3}}+\alpha_{{1,3}}\alpha_{{4,1}}+\alpha_{{1,1}}\alpha_{{2,3}}+\alpha_
{{4,3}}\alpha_{{5,1}}+\alpha_{{2,1}}\alpha_{{3,3}}+\alpha_{{3,1}}
\alpha_{{4,3}}+\alpha_{{1,1}}\alpha_{{4,3}},
\\\nonumber
&& 
-1 +\alpha_{{3,3}}\alpha_{{4,2
}}+\alpha_{{1,2}}\alpha_{{2,3}}+\alpha_{{1,2}}\alpha_{{4,3}}+\alpha_{{
1,3}}\alpha_{{2,2}}+\alpha_{{1,3}}\alpha_{{4,2}}+\alpha_{{2,3}}\alpha_
{{3,2}}+\alpha_{{2,2}}\alpha_{{3,3}}+\alpha_{{4,2}}\alpha_{{5,3}}+
\alpha_{{3,2}}\alpha_{{4,3}}+\alpha_{{4,3}}\alpha_{{5,2}}.  \nonumber
\end{eqnarray}
Finally, the {connected fiber condition} is given by
\begin{eqnarray}\nonumber
&&
-\alpha_{{1,1}}\alpha_{{2,1}}\alpha_{{5,1
}},-\alpha_{{1,1}}\alpha_{{3,1}}\alpha_{{5,1}},-\alpha_{{1,2}}\alpha_{
{2,2}}\alpha_{{5,2}},
-\alpha_{{1,2}}\alpha_{{3,2}}\alpha_{{5,2}},-
\alpha_{{1,3}}\alpha_{{2,3}}\alpha_{{5,3}},-\alpha_{{1,3}}\alpha_{{3,3
}}\alpha_{{5,3}}
\\\nonumber
&&
-\alpha_{{2,1}}\alpha_{{3,1}}\alpha_{{5,1}},-\alpha_{
{2,1}}\alpha_{{4,1}}\alpha_{{5,1}},-\alpha_{{2,2}}\alpha_{{3,2}}\alpha
_{{5,2}},-\alpha_{{2,2}}\alpha_{{4,2}}\alpha_{{5,2}},-\alpha_{{2,3}}
\alpha_{{3,3}}\alpha_{{5,3}},-\alpha_{{2,3}}\alpha_{{4,3}}\alpha_{{5,3
}},
\\\nonumber
&&
\quad
\alpha_{{2,1}}\alpha_{{5,1}},\alpha_{{2,2}}\alpha_{{5,2}},
\alpha_{{2,3}}\alpha_{{5,3}}.
\end{eqnarray}    
}%
The reduced  Groebner basis of the resulted system (computed using the ordering $\alpha \succ  \beta$ ) is given by  
\begin{eqnarray}\label{GBSimpleExple}\nonumber
\mathcal B &=&   \left\{ \beta_{{1}}-1,\beta_{{2}}-1,\beta_{{3}}-1,\beta_{{4}}-1, 
 \beta^2_i-\beta_i,\, \alpha_{ij}^2 -\alpha_{ij}, \,\right.
\\[3mm]\nonumber
&&
\alpha_{{1,2}}\alpha_{{1,3}},\alpha_{{1,2}}\alpha_{{3,2}},
\alpha_{{1,3}}\alpha_{{3,3}},\alpha_{{2,2}}\alpha_{{2,3}},\alpha_{{2,2
}}\alpha_{{4,2}},\alpha_{{2,2}}\alpha_{{5,2}},\alpha_{{2,3}}\alpha_{{4
,3}},\alpha_{{2,3}}\alpha_{{5,3}},\alpha_{{3,2}}\alpha_{{3,3}}, \alpha_{{4,2}}\alpha_{{4,3}},
\\[3mm]\nonumber
&& 
\alpha_{{4,2}}\alpha_{{5,3}},
\alpha_{{4,3}}\alpha_{{5,2}},\alpha_{{5,2}}\alpha_{{5,3}},\alpha_{{4,2}}\alpha_{{5,2
}}-\alpha_{{5,2}},\alpha_{{4,2}}\beta_{{5}}-\alpha_{{5,2}},\alpha_{{4,
3}}\alpha_{{5,3}}-\alpha_{{5,3}}, 
\\\nonumber
&&
\quad \quad \vdots\\\nonumber
&&
-\alpha_{{2,2
}}\alpha_{{5,3}}-\alpha_{{3,2}}\alpha_{{5,3}}+\alpha_{{1,2}}\beta_{{5}
}+\alpha_{{2,2}}\beta_{{5}}+\alpha_{{3,2}}\beta_{{5}}+\alpha_{{3,3}}
\beta_{{5}}+\alpha_{{5,2}}+\alpha_{{5,3}}-\beta_{{5}}\left.\right \}. 
\end{eqnarray}
 In particular, the intersection $\mathcal B\cap \mathbb Q[\beta] =(\beta_{{1}}-1,\beta_{{2}}-1,\beta_{{3}}-1,\beta_{{4}}-1, {\beta_{{5}}}^{2}-\beta_{{5}})$ gives
the two $Y$ minors (i.e., subgraphs $X^\beta$) inside $X$.  
 The remainder of $\mathcal B$ gives
the explicit expressions of the corresponding mappings.
\end{example}

%
%

This feature of Groebner bases can be 
made more precise as follows: 
\begin{theorem}
Let $\mathcal{I}\subset \mathbb Q[x_0, \ldots, x_{n-1}]$ be an ideal, and let $\mathcal{B}$ be a reduced Groebnber basis of~$\mathcal{I}$
with respect to the lex order $x_0\succ \ldots \succ x_{n-1}$. Then, for every $0\leq l\leq n-1$, the set 
\begin{equation}\label{intersectionB}
    \mathcal{B}\cap \mathbb Q[x_{l}, \ldots, x_{n-1}]
\end{equation}
is a  Groebner basis of the ideal $\mathcal{I}\cap Q[x_{l}, \ldots, x_{n-1}]$.
\end{theorem}
Let us conclude by explaining  how the number of solutions of an algebraic systems $\mathcal I\subset \mathbb Q[x_0,\cdots, x_{n-1}]$ can be read from the Groebner basis.   
  This is done using {\it staircase diagrams}, as follows. To each polynomial in $\mathcal I$
we assign a point in the Euclidean space $\mathbb E^{n}$ given by the exponents of its leading term (with respect to the given monomial order). 
The key idea is given by the following proposition: 
 \begin{proposition}
 The ideal $\mathcal I\subset \mathbb Q[x_0,\cdots, x_{n-1}]$ is zero dimensional if and only if 
 the number of points under the shaded region of its staircase is finite, and this number
 is equal to the dimension of the quotient $Q[x_0,\cdots, x_{n-1}]/\mathcal I$, that is, the number of zeros of $\mathcal I$. 
 \end{proposition}
\end{document}